\documentclass[sigconf]{acmart}
\usepackage{multirow, adjustbox}
\usepackage{subcaption}
\usepackage{hyperref}

\AtBeginDocument{%
  \providecommand\BibTeX{{%
    \normalfont B\kern-0.5em{\scshape i\kern-0.25em b}\kern-0.8em\TeX}}}

\setcopyright{acmcopyright}
\copyrightyear{2018}
\acmYear{2018}
\acmDOI{10.1145/1122445.1122456}

\acmConference[Woodstock '18]{Woodstock '18: ACM Symposium on Neural
  Gaze Detection}{June 03--05, 2018}{Woodstock, NY}
\acmBooktitle{Woodstock '18: ACM Symposium on Neural Gaze Detection,
  June 03--05, 2018, Woodstock, NY}
\acmPrice{15.00}
\acmISBN{978-1-4503-XXXX-X/18/06}

\begin{document}

\title{ModelPS: An Interactive and Collaborative Platform for Editing Pre-trained Models at Scale}


\author{Yuanming Li}
\authornote{Both authors contributed equally to this research.}
\affiliation{
    \institution{Nanyang Technological University}
    \country{Singapore}
}
\email{yli056@e.ntu.edu.sg}

\author{Huaizheng Zhang}
\authornotemark[1]
\affiliation{
    \institution{Nanyang Technological University}
    \country{Singapore}
}
\email{huaizhen001@e.ntu.edu.sg}

\author{Shanshan Jiang}
\affiliation{
    \institution{Nanyang Technological University}
    \country{Singapore}
}
\email{univerone@outlook.com}

\author{Fan Yang}
\affiliation{
    \institution{Nanyang Technological University}
    \country{Singapore}
}
\email{yang.fan@ntu.edu.sg}




\author{Yonggang Wen}
\affiliation{
    \institution{Nanyang Technological University}
    \country{Singapore}
}
\email{ygwen@ntu.edu.sg}

\author{Yong Luo}
\affiliation{
    \institution{Wuhan University}
    \city{Wuhan}
    \state{Hubei}
    \country{China}
}
\email{yluo180@gmail.com}





\begin{abstract}


AI engineering has emerged as a crucial discipline to democratize deep neural network (DNN) models among software developers with a diverse background. In particular, altering these DNN models in the deployment stage posits a tremendous challenge. In this research, we propose and develop a low-code solution, ModelPS (an acronym for ``Model Photoshop''), to enable and empower collaborative DNN model editing and intelligent model serving. The ModelPS solution embodies two transformative features: 1) a user-friendly web interface for a developer team to share and edit DNN models pictorially, in a low-code fashion, and 2) a model genie engine in the backend to aid developers in customizing model editing configurations for given deployment requirements or constraints. Our case studies with a wide range of deep learning (DL) models show that the system can tremendously reduce both development and communication overheads with improved productivity.

\end{abstract}





\maketitle

\section{Introduction}
\label{sec:introduction}


Across industries, massive resources have been invested in developing deep neural networks (DNNs)-based applications, ranging from video shopping \cite{hysia}, recipe learning \cite{li2020reciptor}, to resume assessment \cite{luo2019resumegan}. To realize such systems and meet the business requirements, a development team must undergo a series of tasks, including designing, training, optimization, and deployment. Due to the high cost of both money and labor, only 11\% of organizations can deliver a DNN-based product within a week \cite{algorithmia2021enterprise}. To streamline the model development, some platforms, such as Papers With Code \cite{paperswithcode}, are proposed to provide a collection of off-the-shelf pre-trained models \cite{xu2020layoutlm, haldar2019applying} for fast prototype building.

Though these tools alleviated the problem to some degree, they are still far from meeting the efficiency needs resulting from the rapid changes of business requirements. For instance, to build applications with out-of-the-box models, developers first need to learn the structure of these models and their usage by understanding the source code. Then, they need to try out multiple plausible strategies, such as selecting transfer learning (TL) methods with different data augmentations, to eventually furnish models that can meet their requirements. The process is usually filled with tedious and costly tasks, such as writing codes and testing model performance \cite{zhang2020no} repetitively.



Several attempts have been made to further improve modeling productivity, as summarized in Table \ref{tab:system_compare}. These approaches fall into three categories, visualization tools, drag-and-drop tools, and TL encapsulation tools. The first type, such as \cite{netron}, reduces user effort by providing intuitive model structure visualization, speeding up the learning process. The second approach allows users to build DNN models from scratch by dragging and dropping basic blocks, such as a convolution layer \cite{ennui}. The third type abstracts and encapsulates many TL algorithms to offer expert users out-of-the-box TL tools with higher efficiency \cite{qiu2020easytransfer, tlt}. 

\begin{table*}[ht]
\adjustbox{max width=\textwidth}{%
\begin{tabular}{lccccccc}
\hline
Name                                    & Visualizer & Model Editor & Model Export & TL Training & Feature Selector & Model History & Workspace  Sharing \\ \hline
Netron\cite{netron}                     & \checkmark &              &              &             &                  &                           &                               \\ \hline
ENNUI\cite{ennui}                       & \checkmark &              & \checkmark   &             &                  &                           &                               \\ \hline
EasyTransfer\cite{qiu2020easytransfer}  &            &              &              & \checkmark  &                  &                           &                               \\ \hline
TLT\cite{tlt}                           &            &              &              & \checkmark  & \checkmark       &                           &                                \\ \hline
\textbf{ModelPS}                        & \checkmark & \checkmark   & \checkmark   & \checkmark  & \checkmark       & \checkmark                & \checkmark                     \\ \hline
\end{tabular}%
}
\caption{Comparison of competing platforms. ModelPS provides a complete solution for teams to edit and improve ML models collaboratively and interactively.}
\label{tab:system_compare}
\end{table*}

Despite the considerable progress, problems remain with these systems. First, they do not provide an end-to-end integrated workflow for users to easily share, learn, and edit DNNs. Instead, they treat these steps separately, which saves very little user effort — Users are nonetheless enforced to check the model usage manual and write boilerplate codes. Second, it is difficult for non-expert users to utilize the current out-of-the-box TL tools provided in these approaches which require domain knowledge from users, including algorithms and training hyper-parameters. Third, developing a DNN application is often teamwork, but the existing solutions do not offer collaborative tools to further reduce the communication cost.

To narrow the gaps, we embarked on designing a holistic system guided by the following principles. First, the system should provide an interactive and intuitive interface to support the entire model improvement process, from model visualization, edition, validation, to training. Second, the system should consider both expert and non-expert users, enabling them to modify their DNN applications with ease. Third, the system should facilitate team collaborations to speed up the development process. In addition, the system should be easily extended to support new use cases and absorb ever-emerging DNN tools.

Under these guidelines, we have designed a new system, termed ModelPS, providing an interactive and collaborative solution for pre-trained model modification. It contains an intuitive web interface and a comprehensive backend server. The former allows users to publish, visualize, and modify DNNs and explore the DNN performance under different edit configurations. Users can enjoy the experience of `what you see is what you get'. At the same time, the backend server provides a set of modules for operation, including a TL zoo and feature store, which can be easily customized and extended. We also developed a Model Genie to streamline the process of setting screening for achieving product goals. In addition, the whole system is designed for teamwork by providing rich collaboration tools. The following case studies demonstrated the system's usability, effectiveness, and efficiency.
 

\section{System Highlights}

In this section, we first highlight the key features of ModelPS and then show the uniqueness of the platform through a side-by-side comparison with competing products.

\subsection{Features}

The system is characterized by the following five features.

\textbf{Interactive}. As a low-code platform, ModelPS ensures that both researchers and practitioners do not need to write any boilerplate code for model modification. It further offers several preview modules (e.g., feature preview) and fast-validation functions so that users can intuitively fine-tune a model like how designers use PhotoShop.

\textbf{Holistic}. ModelPS provides a one-stop service, encapsulating all tasks associated with model modification for a team. Besides editing, the system supports model publishing and sharing, as well as housekeeping. It also supports data and TL management to further speed up the modification process.

\textbf{Collaborative}. Our design improves team productivity by providing collaborative space for smooth teamwork. The system offers various user roles in a team. Model designers can agilely develop their DNN frameworks while DevOps engineers can conduct performance evaluations against product requirements. Data engineers can extend the system by designing new data augmentation methods, and data scientists can construct new TL methods with the provided APIs. 

\textbf{Intelligent}. We design and incorporate intelligent modules, such as Model Genie, to assist with DNN customization for non-experts. The system also automates the entire workflow, including training task encapsulation and job dispatch to accelerate the process and improve resource utilization.

\textbf{Extensive.} The modular design of ModelPS enables further customized extensions, such as incorporating new datasets and state-of-the-art TL algorithms, as well as docking into other training and serving systems.


\subsection{Competitive Assessment}


We summarize competing toolkits in Table \ref{tab:system_compare}. ModelPS differs from these systems in three aspects:

(1) Other systems do not adhere to the new pre-training paradigm, while our framework is designed for modifying pre-trained models and transforming them into products. For instance, the system developed by ENNUI \cite{ennui} focuses on building models from scratch in a drag-and-drop fashion. In contrast, our system is equipped with a pre-trained model editing function inherently.

(2) Other systems treat each editing stage individually, while our solution provides end-to-end workflow with many supporting functions. For example, the system proposed by Neutron \cite{netron} only offers model visualization, while other tools such as \cite{qiu2020easytransfer, tlt} merely provide TL algorithms for experiments. Without an integrated workflow, these systems often leave users with tedious tasks, including code replication and modification sync, hindering further productivity improvement.

(3) As a lightweight system, ModelPS is highly customizable. It can work efficiently with popular frameworks, such as training system PyTorch Lighting \cite{falcon2019pytorch}, model deployment platform Clipper \cite{crankshaw2017clipper} and MLModelCI \cite{zhang2020mlmodelci}. Those other platforms capable of pre-trained model editing, such as \cite{nnc, dls}, are closed source and therefore can not be quickly extended to upgrade fast-evolving DNNs. Besides, all these systems demand users to accommodate their built-in toolchains, introducing unpredictable risks and sharp learning curves, often with expensive subscription fees.


\section{System Architecture}

This section provides an overview of the ModelPS architecture, including the system workflow and the detailed descriptions of the core components. 


\subsection{Workflow}
\label{sec:workflow}

\begin{figure}[t]
  \centering
  \includegraphics[width=\linewidth]{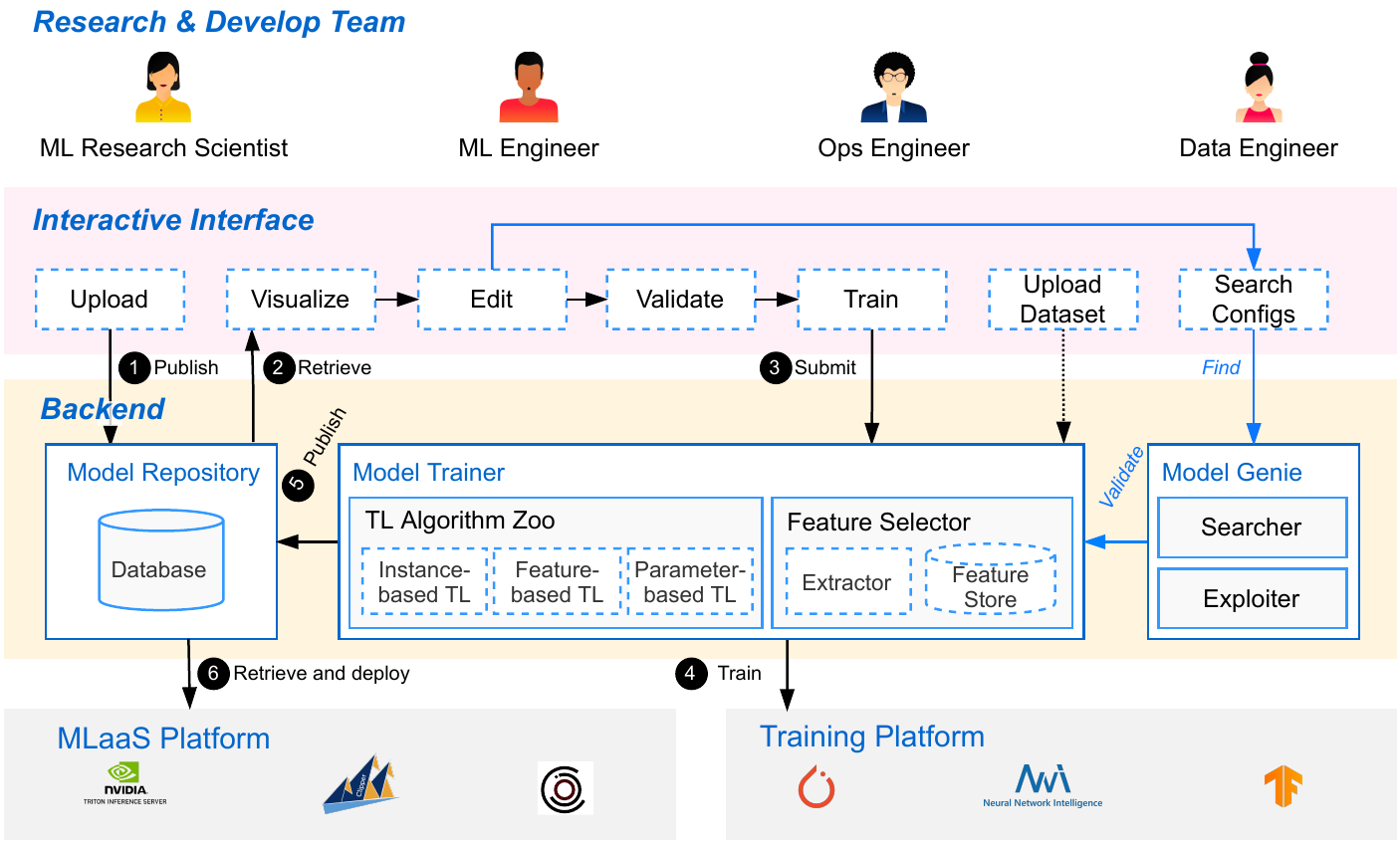}
  \caption{ModelPS overview. The system consists of an interactive interface and an intelligent backend for team members to improve their ML services. It works as a bridge that seamlessly connects existing training and deployment systems, streamlining the ML product development.}
  \label{fig:workflow}
  \Description{modelps\_workflow}
\end{figure}

Figure \ref{fig:workflow} depicts the workflow for a team to modify DNN models in ModelPS. First, Model designers upload their works to a centralized model repository for team members to access. A visualizer is provided for easy understanding of these models. Within the same page, users can interactively edit the model structures and create new ones. During the process, various options, such as different TL methods, datasets, and feature extractors, are provided to meet the requirements in terms of accuracy, latency, etc. Also, we offer a validator for quick verification of these modifications. To simplify the configuration for non-experts, we implement a Model Genie to facilitate automatically. Upon the completion of modification and verification, ModelPS will parse and encapsulate the user settings into the model training job which will be completed with cluster resources. After training is finished, the system will automatically deposit the updated model in the repository and show it to users in a richly designed dashboard.

\subsection{Model Repository}
\label{model-hub}

ModelPS relies on a centralized repository for model sharing and housekeeping. The module is built atop MongoDB and abstracts models on two levels: the model weight file and the meta information such as accuracy and parent model ID. It primarily provides three APIs: (1) \texttt{publish} accepts a model weight file along with its metadata from model designers; (2) \texttt{retrieve} searches and caches a model copy locally with the provided query inputs; (3) \texttt{save\_draft} saves unfinalized model structure for future editing or accessing. These APIs pivot around the pre-trained model edit, and can be extended further.

\subsection{Interactive Interface}
\label{interactive-interface}

ModelPS provides an interactive web interface that enables users to view and improve existing models with no coding effort. It consists of the following core components:

\textbf{Visualizer}
\label{visualizer}
displays DNNs in detail, embodying each structural block or layer. The displayed information includes the block name (e.g., residual block), the layer type (e.g., a convolution layer), input dimension, etc. Also, we provide their connection information to indicate the data flow direction. We further provide the model overview for reference, including a pre-trained dataset, and model performance.

\textbf{Editor}
\label{editor}
enables code-free model modification, which is abstracted on two levels, the layer and the ensemble. For instance, right-clicking the layer icon will pop up an interactive panel to guide users to adjust the layer configuration to meet their requirements. Meanwhile, users can switch between different pre-trained models, datasets, and data augmentation methods, as well as adjust hyper-parameters, all of which are enveloped under 'ensemble edit'. The editor periodically packs the modification as a serialized JSON string, which can be translated to represent a model draft and saved into the repository. Such a draft can be revised later or opened remotely by other users, through design review and team collaboration.

\textbf{Validator}
\label{sec:validator}
provides a fast quality validation of the modified model. In particular, the validator batches several training samples based on the user configuration and conducts time-limited (i.e., in a few minutes) training, usually in a spare GPU, for the evaluation of accuracy, training cost, and inference latency. The results are stored in the database and a report is displayed on the screen. The process is entirely code-free.

\subsection{Model Trainer}
\label{sec:trainer}
 
A successful model modification heavily relies on the TL training methods (e.g., algorithm, optimizer, and hyper-parameters) and data quality (e.g., data cleaning and data argumentation), which can introduce multi-dimensional complexity to the training task. Therefore, a model trainer is implemented to shield such complexity from users and enable them to utilize models with ease. The module consists of two functional bodies, TL zoo and feature selector, and controls all the training processes. To facilitate users to manage the process in a cluster and customize the module, the system also provides a set of APIs, including \texttt{start}, \texttt{terminate}, \texttt{pause} and \texttt{to\_device}. We implement the module based on the PyTorch Lighting, and users can dock the system to existing training and other data processing systems.

\textbf{TL Algorithm Zoo}
\label{sec:tl-algorithm-zoo} 
includes a set of TL algorithms to streamline the editing of pre-trained models for new tasks. These methods can be categorized into instance-based TL (Tradaboost \cite{pardoe2010boosting}), feature-based TL (Deep Adaption Network \cite{long2015learning}), and parameter-based TL (fine-tuning, knowledge distillation \cite{hinton2015distilling}) as discussed in \cite{pan2009survey}, and researchers can build their solutions by inheriting and customizing these algorithms.

\textbf{Feature Selector}
\label{sec:feature-selector} 
consists of two functional components, feature store and feature extractor. The feature store enables users to publish and index datasets (which can be either original data or extracted features). Authorized users are allowed to access these datasets for model editing. The feature extractor contains many build-in data argumentation methods and provides an out-of-the-box user experience. These methodological options empower users with more possibilities for achieving better model performance.

\begin{figure}
    \centering
    \includegraphics[width=1.0\linewidth]{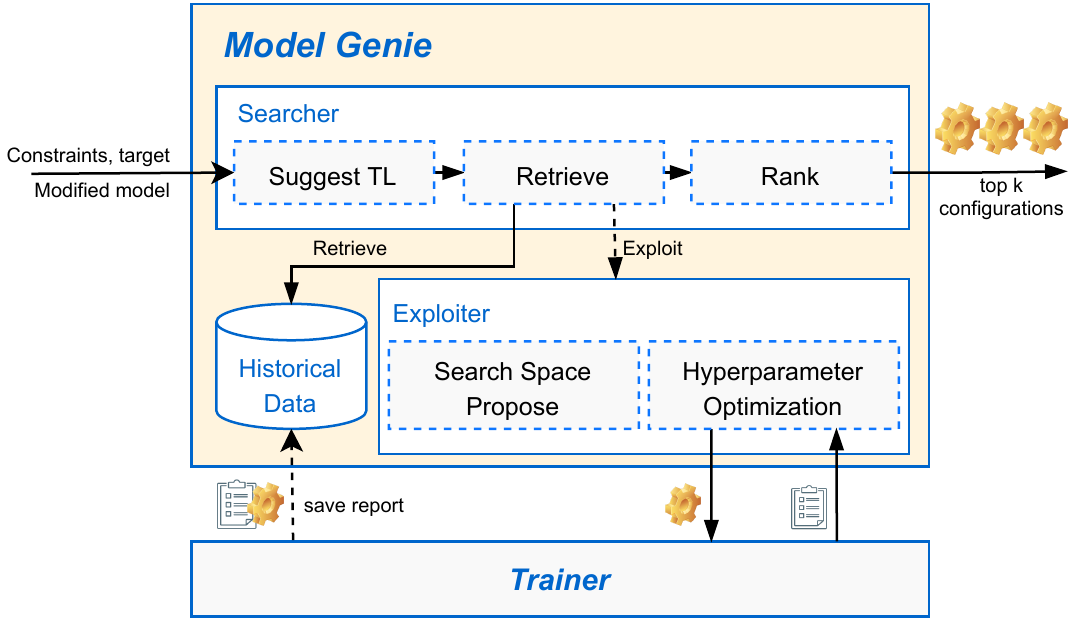}
    \caption{Model Genie configuration search workflow. The module suggests suitable results from historical data. It will explore more training configurations in the case of insufficient search results are presented.}
    \label{fig:model_genie_workflow}
\end{figure}

\subsection{Model Genie}
\label{sec:model_genie}
It is a non-trivial task to select an optimal TL training configuration from many optional combinations to satisfy specific constraints (e.g., latency Service Level Objective (SLO)) and targets (e.g., higher accuracy). This motivates us to develop Model Genie, which automatically suggests optimal configuration for users as shown in Figure \ref{fig:model_genie_workflow}. It consists of two sub-modules, a searcher and an exploiter.

\textbf{Searcher}
translates user specifications (i.e., constraints $\textbf{C} = \{c_i\}$ and targets $\textbf{T} = \{t_i\}$) into a query to the historical TL training data $\textbf{R}$. The query result $\textbf{Q}$ is described by relational algebra as $\textbf{Q} \coloneqq \tau_{\textbf{T}}({\sigma_{c_i \in \textbf{C}, tl}(\textbf{R})})$, where the recommended TL method $tl$ is generated from user-specified constraints $\textbf{C}$ based on pre-defined rules, $\sigma_{c}(\textbf{R})$ is the selection of records in $\textbf{R}$ holding conditions $c$, and $\tau_{A}(\textbf{R})$ describes the sorting operation in records $\textbf{R}$ according to attribute $A$. The top results of $\textbf{Q}$ will be recommended as preferred configurations.

\textbf{Exploiter}
takes two steps to find suitable configurations in the case of insufficient historical training configurations. First, the exploiter proposes a search space in a rule-based approach, with a specified model, TL method, and target dataset. The search space contains possible pre-trained model structures, pre-trained datasets, data augmentations, and training hyperparameters (e.g., learning rate and the number of fine-tuning layers). Next, the built-in hyperparameter optimization (HPO) method iteratively selects configuration, one at a time, for the trainer to evaluate. The validation report, containing performance information such as accuracy and inference latency, will guide the selection of trial configuration for the next iteration.

\begin{figure}[!ht]
\centering

    \begin{subfigure}{0.48\textwidth}
        \centering
        \includegraphics[width=1.0\linewidth]{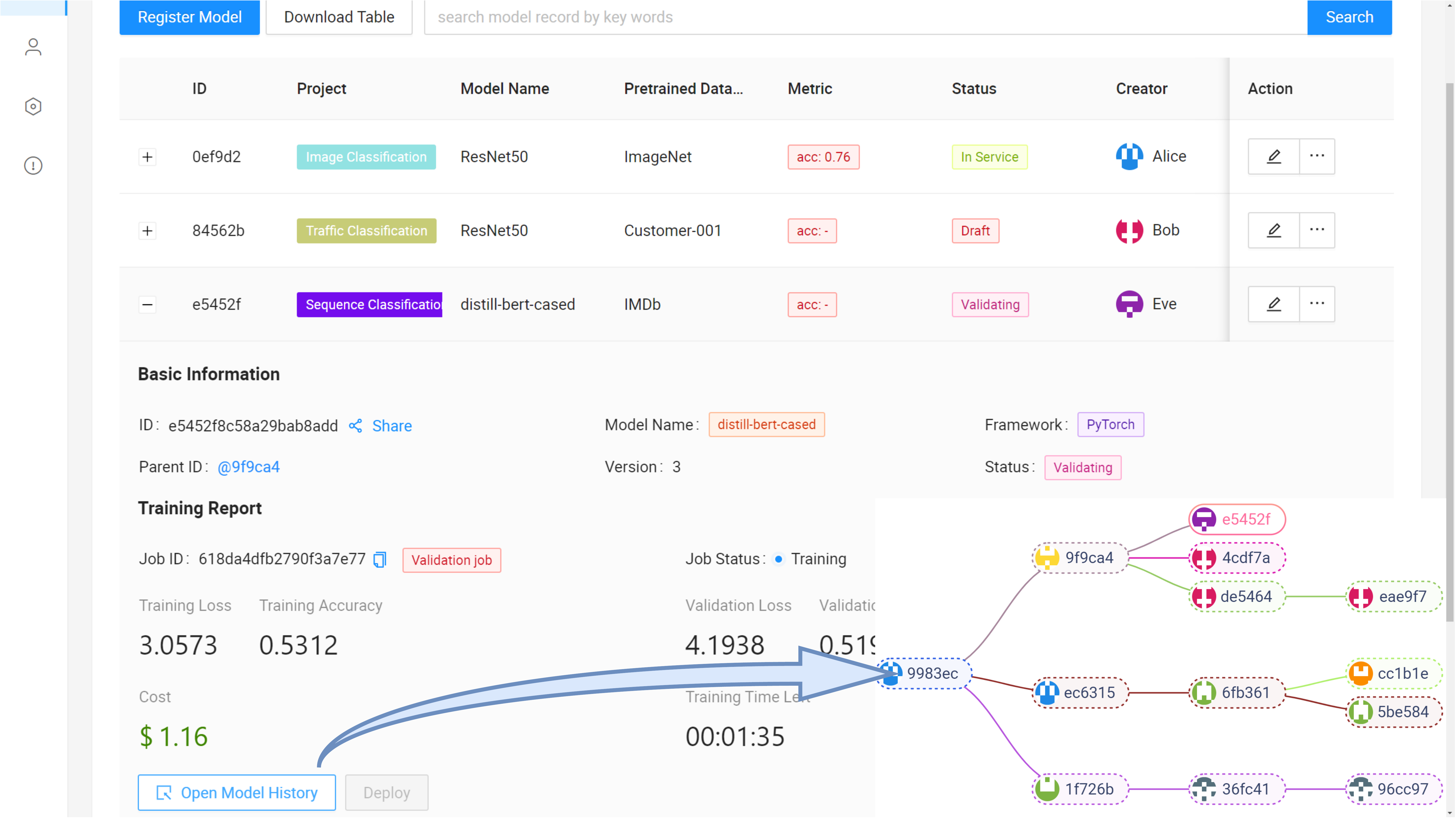}
        \caption{ModelPS Dashboard}
        \label{fig:model_dashboard}
    \end{subfigure}\hfill%
    \begin{subfigure}{0.48\textwidth}
        \centering
        \includegraphics[width=1.0\linewidth]{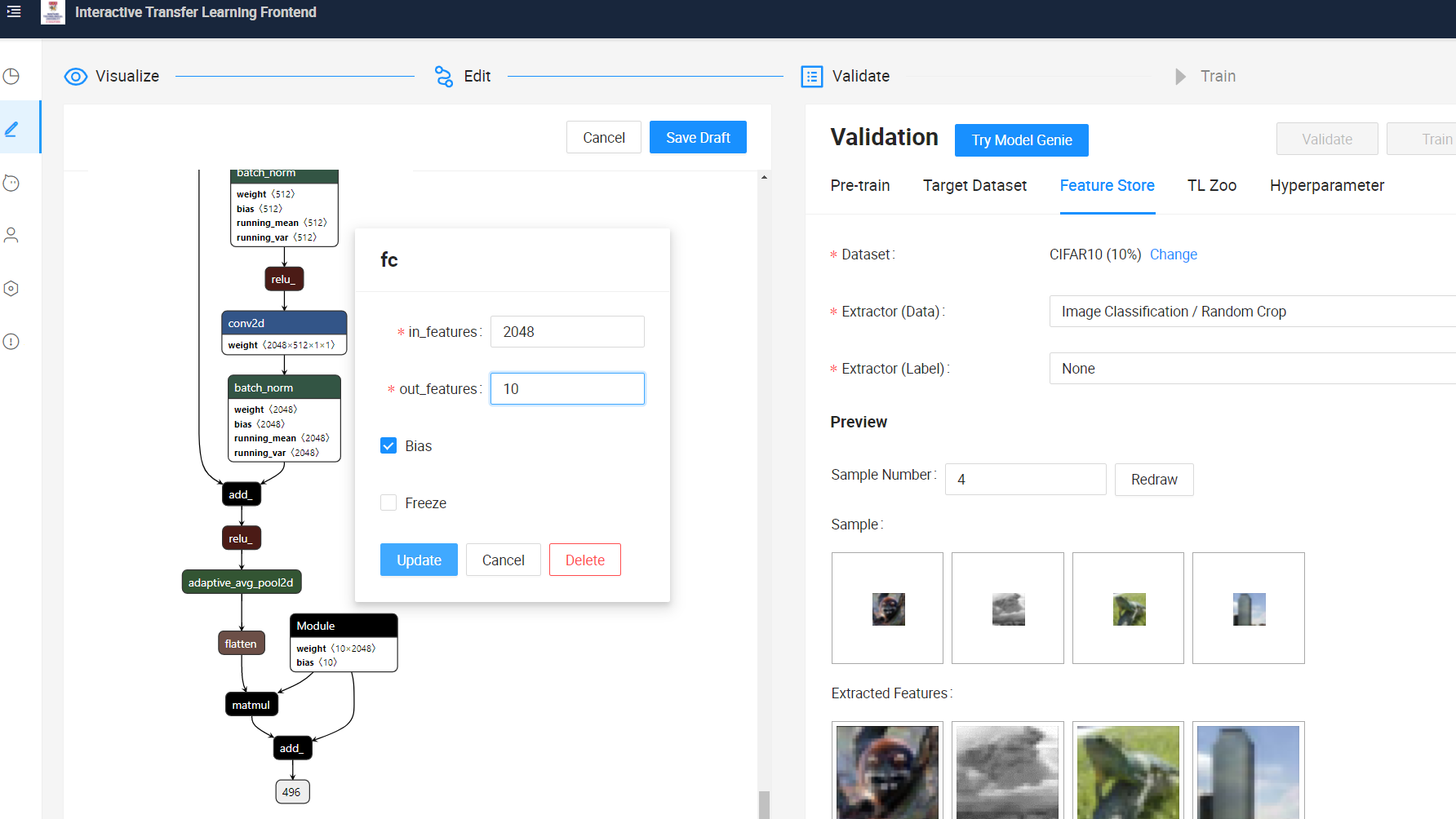}
        \caption{Interactive Editor}
        \label{fig:model_editor}
    \end{subfigure}\hfill%

\caption{Two main interactive web interfaces. They show a complete and intuitive model edit workflow through a model management module (Fig.\ref{fig:model_dashboard}) and an interactive model editor (Fig.\ref{fig:model_editor}). Fig.\ref{fig:model_dashboard} shows that the system organizes many models uploaded by model designers, where team players can check their details, such as model edit history. Fig.\ref{fig:model_editor} illustrates the edit and validation interface where users need to consider many options (e.g., pre-trained model selection, feature engineering, and deployment scenarios) that may influence final results.}
\label{fig:model_ps_frontend}
\end{figure}

\section{Demonstration}
\label{sec:demostration}

In this section, we present three case studies\footnote{Check demonstration at \url{https://cap-ntu.github.io/modelps_demo}} to demonstrate that ModelPS is able to fulfill the following requirements (RQ) for an R\&D (research \& develop) team:\\
\textbf{RQ1}: a no-code interface for swiftly editing and improving pre-trained models using diverse TL algorithms.\\
\textbf{RQ2}: a Model Genie to streamline the model edit process.\\
\textbf{RQ3}: a shared workspace that facilitates team collaboration to reduce the administration and communication cost during the ML product development.

\begin{figure}[!ht]
\centering

    \begin{subfigure}{0.24\textwidth}
        \centering
        \includegraphics[width=1.0\linewidth]{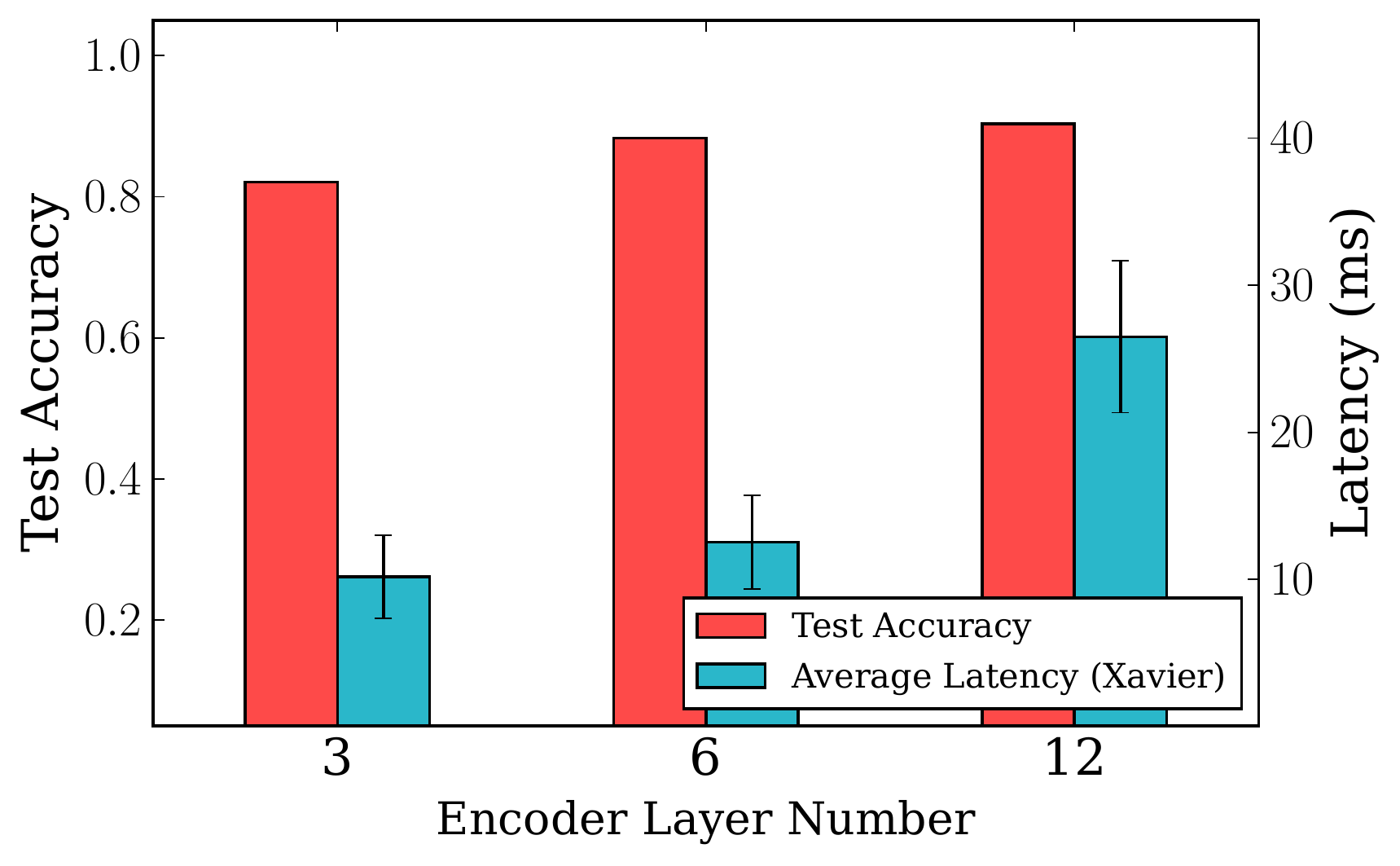}
        \caption{Pre-trained Model}
        \label{fig:pre_trained_model}
    \end{subfigure}\hfill%
    \begin{subfigure}{0.22\textwidth}
        \centering
        \includegraphics[width=1.0\linewidth]{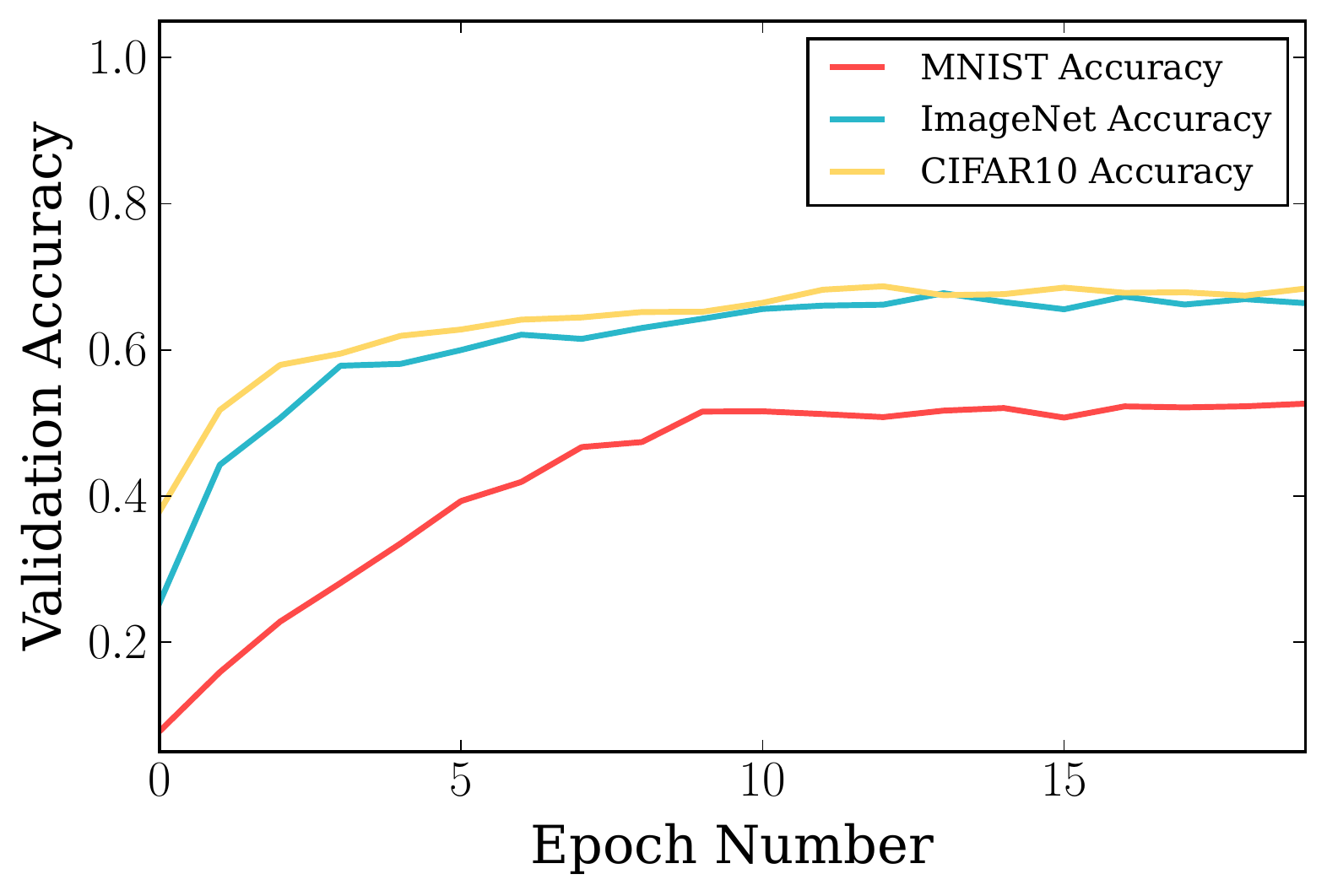}
        \caption{Pre-trained Dataset}
        \label{fig:pretrained_dataset}
    \end{subfigure}%

    \begin{subfigure}{0.24\textwidth}
        \centering
        \includegraphics[width=1.0\linewidth]{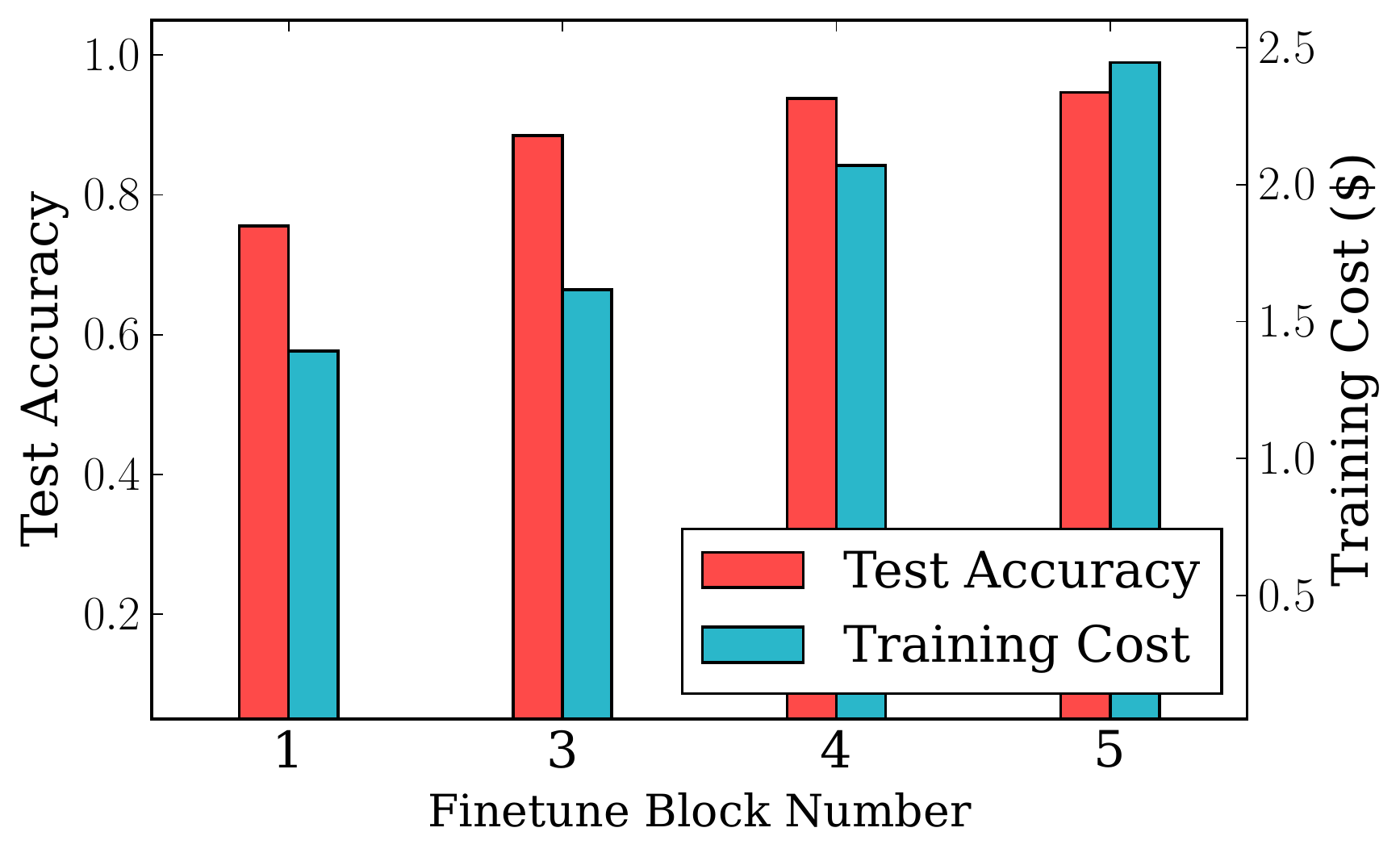}
        \caption{TL Hyperparameters}
        \label{fig:inference_training_time_tl_hyperparameter}
    \end{subfigure}%
        \begin{subfigure}{0.22\textwidth}
        \centering
        \includegraphics[width=1.0\linewidth]{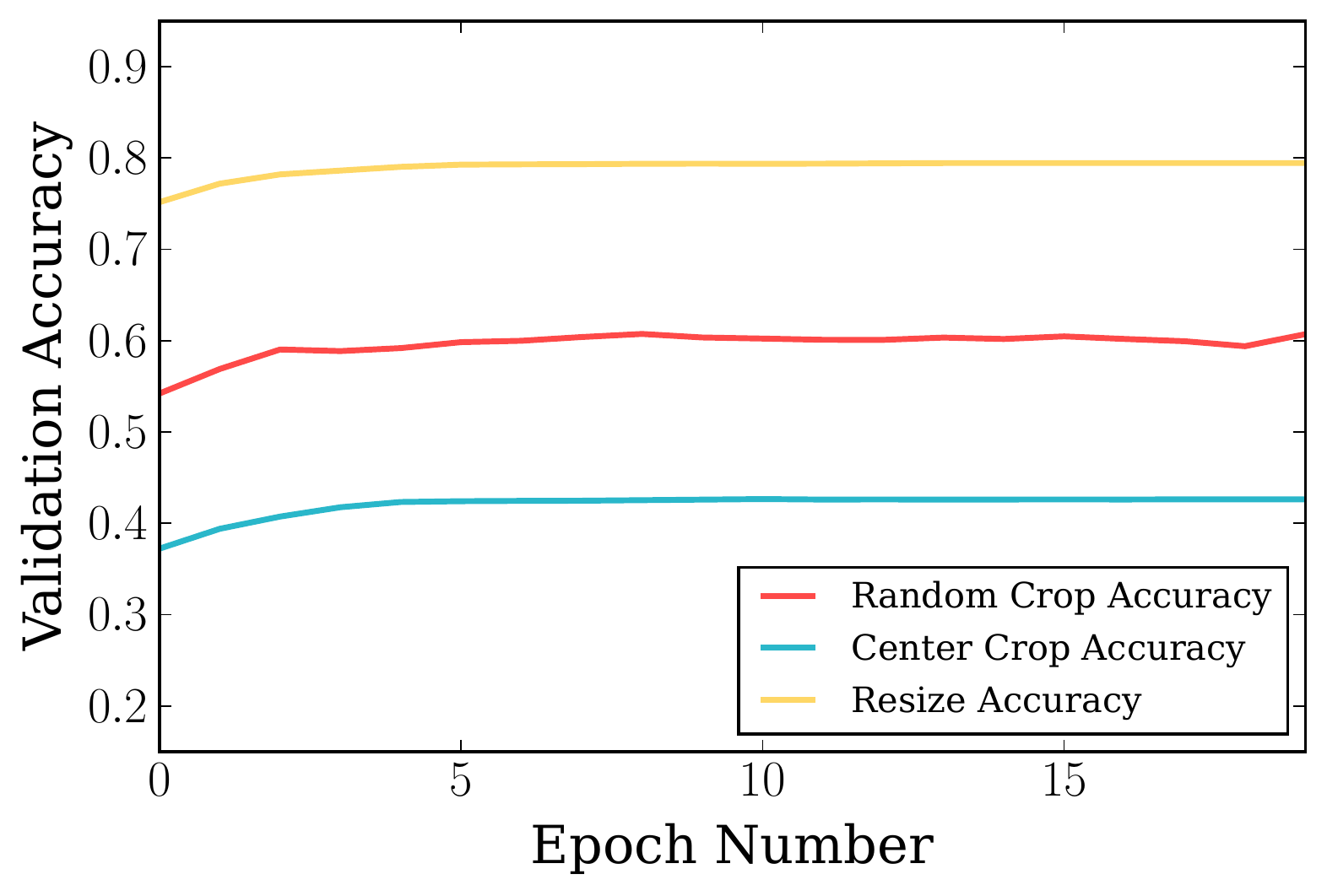}
        \caption{Feature Selection}
        \label{fig:feature_selection}
    \end{subfigure}%

\caption{Different configurations affect model accuracy, latency, and training cost.}
\label{fig:model_ps_study}
\end{figure}

\textbf{RQ1.} Suppose that Alice is a data scientist (DS) with expertise in the traffic analysis, and she plans to build a DL-based street view classification application. To verify her idea quickly, she decides to make use of TL to adjust the pre-trained DL models of image classification (IC), designed by her teammates. Through ModelPS, Alice can achieve this goal without writing any script. First, she turns to the web interface and screens the related models stored in the system, as shown in Figure \ref{fig:model_dashboard}. Then, she selects the ResNet50 model pre-trained on ImageNet to start her tuning journey. The model structure will be visualized on a new page, as shown in Figure \ref{fig:model_editor}. Alice can right-click the layers and tune their parameters in a pop-up window. Subsequently, given the high-accuracy requirement, Alice tries many different options to improve the model performance. As shown in Figure \ref{fig:model_ps_study}, these methods include a) Switching between different pre-trained models without much manual intervention; b) Adopting a pre-trained dataset more similar to the target one; c) Selecting different hyperparameters (e.g., number of fine-tuning layers) for a specific TL method; d) Performing feature selection. All these actions can be executed on the same page where Alice can preview (e.g., feature type) and validate their configurations timely. Once the adjustment is done, Alice submits her modified model which will be trained and saved in the system automatically.

\begin{figure}[t]
    \centering
    \includegraphics[width=1.0\linewidth]{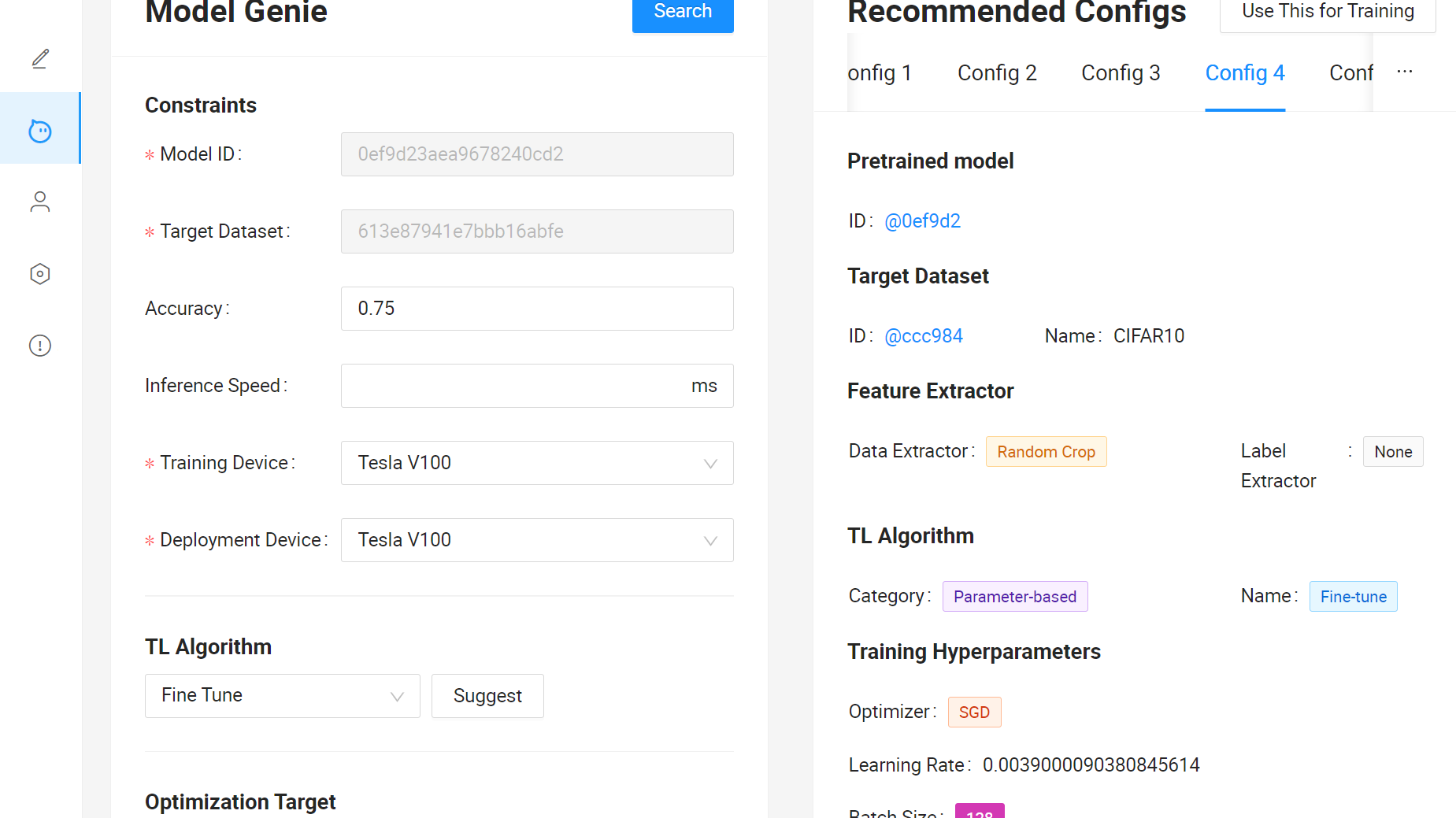}
    \caption{Model Genie helps reduce human efforts on model editing. It only requires minimum constraint input from users to recommend optimal edit configuration.}
    \label{fig:model_genie}
\end{figure}

\begin{figure}[!ht]
\centering

    \begin{subfigure}{0.235\textwidth}
        \centering
        \includegraphics[width=1.0\linewidth]{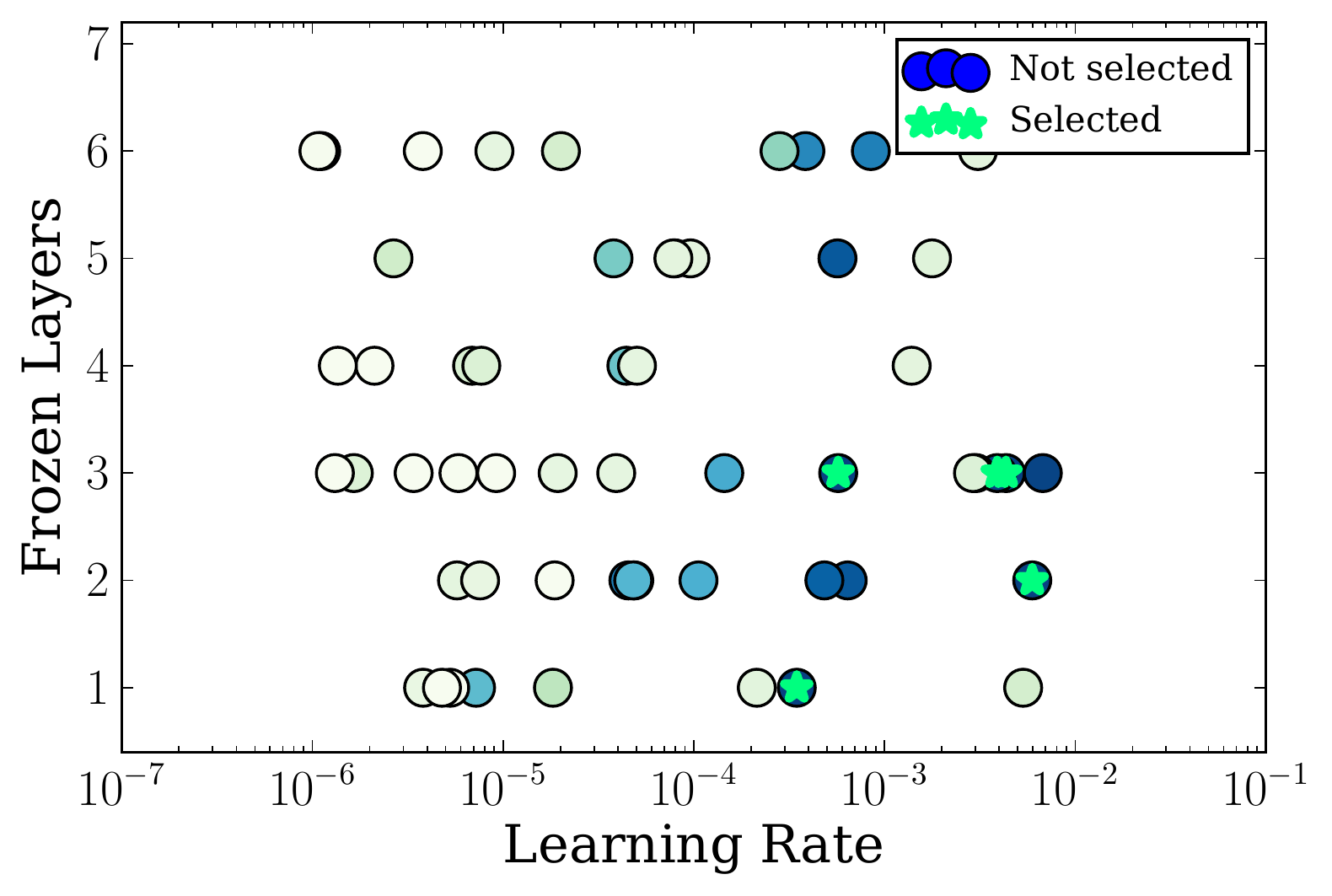}
        \caption{Higher Accuracy}
        \label{fig:model_genie_best_accuracy_search}
    \end{subfigure}\hfill%
    \begin{subfigure}{0.235\textwidth}
        \centering
        \includegraphics[width=1.0\linewidth]{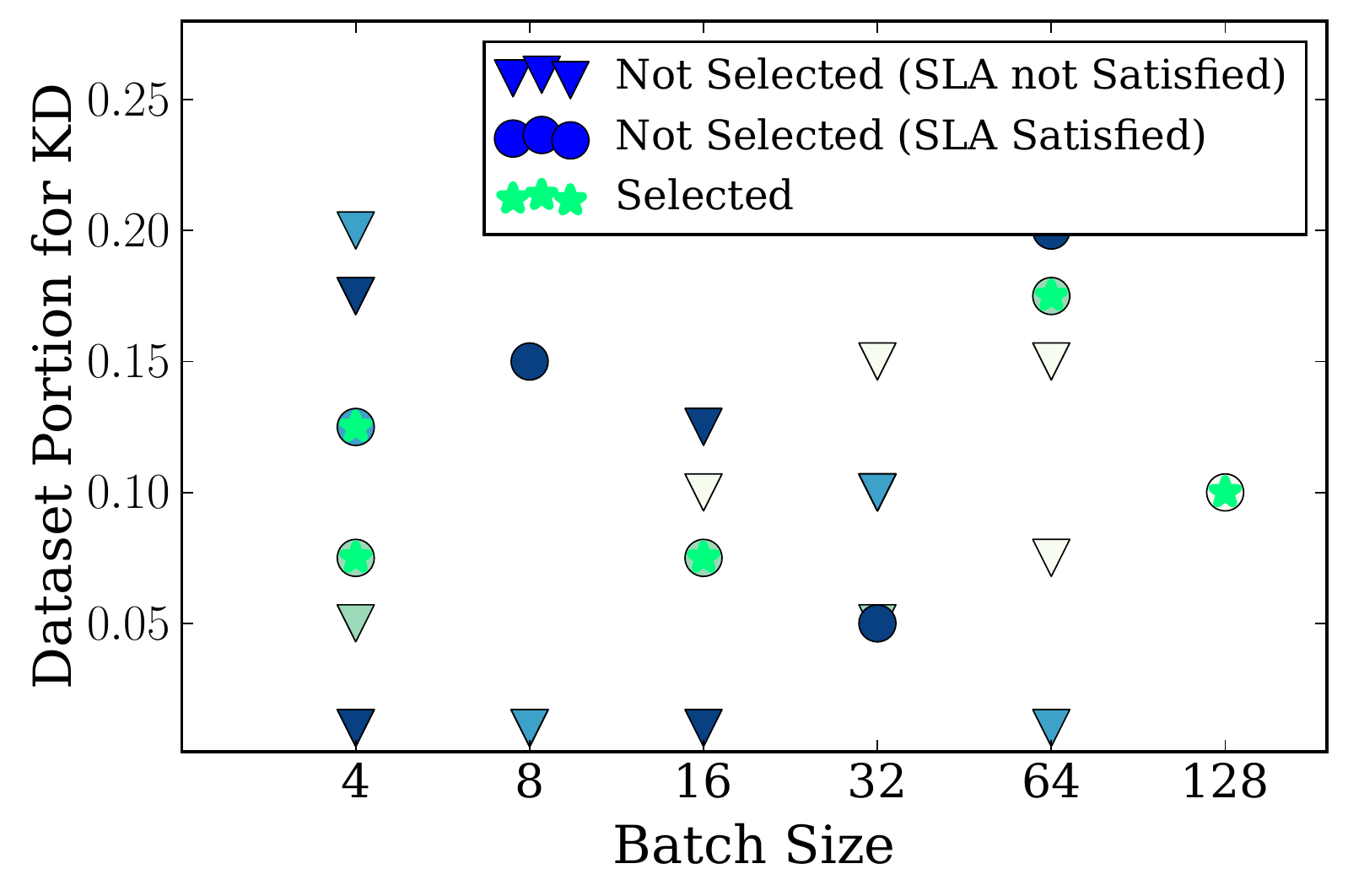}
        \caption{Lower Latency}
        \label{fig:model_genie_fastest_speed_search}
    \end{subfigure}%

\caption{Model Genie searches for configurations with higher accuracy (darker color in Figure \ref{fig:model_genie_best_accuracy_search}) or lower latency (darker color in Figure \ref{fig:model_genie_fastest_speed_search}) using hyperparameter optimization and selects top-5 results (marked as green star).}
\label{fig:model_genie_hpo}
\end{figure}

\textbf{RQ2}. The above case study has shown that there are many combinations of configuration options on the table. This motivates us to design a Model Genie that can aid users, especially non-experts, to perform their tasks. Model Genie considers a joint selection of pre-trained models, TL parameters, and training hyperparameters (e.g., learning rate and data argumentation), given user requirements, such as target latency and accuracy. We illustrate the workflow (Figure \ref{fig:model_genie_workflow}) in two scenarios.

\textbf{Deploying IC Models to achieve higher accuracy on cloud}. In this scenario, all that Alice needs to do is providing the target accuracy. Model Genie will first screen the historical configurations and obtain items with higher accuracy than the target value, and rank such settings by latency. If the candidate configurations are insufficient (e.g., less than 5), Model Genie will explore new settings in parallel using build-in HPO methods. These methods will identify a set of options to fine-tune the model, as shown in Figure  \ref{fig:model_genie_best_accuracy_search}. The result of each configuration will be recorded as historical data to guide other user model improvement processes in the future.

\textbf{Deploying TC models to achieve lower latency on edge}. Suppose that Bob wants to build a real-time TC (text classification) application running on the edge to protect data privacy. Model Genie offers a two-stage approach to achieve his goal. After Bob enters his targets into our system, Model Genie will first perform knowledge distillation (KD) to generate small models with public data on the cloud. Then the models will be dispatched to edge devices and fine-tuned on the collected sensitive dataset. The final results that meet the latency target will be sorted by to accuracy in ascending order, and the top 5 results will be returned for the recommendation. Similarly, Model Genie combines both configuration search and HPO methods. The results are shown in Figure \ref{fig:model_genie_fastest_speed_search}.


\textbf{RQ3}. Developing a highly efficient ML product is teamwork rather than a private battle. Besides the DS (e.g., Alice or Bob), other team members, such as Product Engineer Tom, Data Engineer Mary, and DevOps Engineer Steven, can also leverage the system to make contributions, at a low communication cost. For instance, Tom can obtain the newly trained model performance in Figure \ref{fig:model_dashboard} and further improve the model with other tools (e.g., TVM). Mary can design and feed new datasets or data processing strategies to our system and preview their effects (Figure \ref{fig:model_dashboard}). Steven can check the performance in terms of accuracy, latency, etc., which can offer efficient debugging. Based on the comprehensive results, senior leaders can make the next decisions, such as adding training data and designing better models accordingly. In conclusion, our system offers a highly productive and collaborative working experience to meet DNN editing demands.


\section{Future Research Directions and Discussion}

This section presents several future research directions and discusses their potential challenges and solutions.

\textbf{Joint search of pre-trained models, TL algorithms, and hyper-parameters}. To date, both academia and industry have developed many hyper-parameter optimization (HPO) methods, significantly reducing both human efforts and searching time \cite{snoek2012practical}. Meanwhile, efficiently selecting suitable pre-trained models for specific tasks are being researched on \cite{you_logme:_2021}. However, how to jointly search for them to further reduce human resources and time costs is waiting to be inspected. Our preliminary studies show that searching them together results in a vast search space. Additionally, optimizing the search that involves both categorical parameters (e.g., models and algorithms) and real-valued parameters (e.g., learning rate) remains challenging.

\textbf{Reinforcement learning for scheduling model editing and inference jobs}. The second direction is to investigate reinforcement learning (RL) methods for scheduling model editing and inference jobs in one cluster. The main challenge in this topic is two levels of heterogeneity: job heterogeneity (JH) and resource heterogeneity (RH). First, inference jobs emphasize latency, while fine-tuning jobs care about throughput. The scheduling shall consider specific targets of different jobs separately. Second, ML clusters always consist of machines with enormous differences in computational performance, due to their hardware types: CPUs, GPUs, and FPGAs. Moreover, different architectures in the GPUs (e.g., Pascal and Volta) makes job scheduling in such an environment even harder. In conclusion, designing a rule-based method is infeasible to our problem, and we hope to solve it by applying RL. 

\section{Conclusion}

In this article, we present ModelPS, an interactive and collaborative low-code platform for fast model improvement by a team. ModelPS enables developers and researchers to build and modify models without much domain knowledge on model structures and DL frameworks. It requires minimum manual coding effort and lowers team communication costs. To streamline the model improvement process, the system provides a number of ready-to-use modules, such as model repository, visualizer, editor, validator, and trainer. We also have a heuristic Model Genie, that automatically recommends optimal training configurations for given user requirements. We demonstrate the interactive, collaborative and extensive character of our system, which collectively help users to rapidly create new models for different AI products in various scenarios. We will continually upgrade ModelPS to meet the ever-evolving model editing needs across industry and academia. 

\bibliographystyle{ACM-Reference-Format}
\bibliography{sample-base}


\begin{thebibliography}{23}


\ifx \showCODEN    \undefined \def \showCODEN     #1{\unskip}     \fi
\ifx \showDOI      \undefined \def \showDOI       #1{#1}\fi
\ifx \showISBNx    \undefined \def \showISBNx     #1{\unskip}     \fi
\ifx \showISBNxiii \undefined \def \showISBNxiii  #1{\unskip}     \fi
\ifx \showISSN     \undefined \def \showISSN      #1{\unskip}     \fi
\ifx \showLCCN     \undefined \def \showLCCN      #1{\unskip}     \fi
\ifx \shownote     \undefined \def \shownote      #1{#1}          \fi
\ifx \showarticletitle \undefined \def \showarticletitle #1{#1}   \fi
\ifx \showURL      \undefined \def \showURL       {\relax}        \fi
\providecommand\bibfield[2]{#2}
\providecommand\bibinfo[2]{#2}
\providecommand\natexlab[1]{#1}
\providecommand\showeprint[2][]{arXiv:#2}

\bibitem[\protect\citeauthoryear{Algorithmia}{Algorithmia}{2020}]%
        {algorithmia2021enterprise}
\bibfield{author}{\bibinfo{person}{Algorithmia}.}
  \bibinfo{year}{2020}\natexlab{}.
\newblock \bibinfo{booktitle}{\emph{2021 enterprise trends in machine
  learning}}.
\newblock \bibinfo{publisher}{Algorithmia}. 23 pages.
\newblock
\urldef\tempurl%
\url{https://info.algorithmia.com/hubfs/2020/Reports/2021-Trends-in-ML/Algorithmia_2021_enterprise_ML_trends.pdf?hsLang=en-us}
\showURL{%
\tempurl}
\newblock
\shownote{Accessed: 2021-05-01.}


\bibitem[\protect\citeauthoryear{Cognition}{Cognition}{2020}]%
        {dls}
\bibfield{author}{\bibinfo{person}{Deep Cognition}.}
  \bibinfo{year}{2020}\natexlab{}.
\newblock \bibinfo{title}{Deep Learning Studio}.
\newblock
\newblock
\urldef\tempurl%
\url{https://deepcognition.ai/}
\showURL{%
\tempurl}
\newblock
\shownote{Accessed: 2021-02-26.}


\bibitem[\protect\citeauthoryear{Crankshaw, Wang, Zhou, Franklin, Gonzalez, and
  Stoica}{Crankshaw et~al\mbox{.}}{2017}]%
        {crankshaw2017clipper}
\bibfield{author}{\bibinfo{person}{Daniel Crankshaw}, \bibinfo{person}{Xin
  Wang}, \bibinfo{person}{Guilio Zhou}, \bibinfo{person}{Michael~J Franklin},
  \bibinfo{person}{Joseph~E Gonzalez}, {and} \bibinfo{person}{Ion Stoica}.}
  \bibinfo{year}{2017}\natexlab{}.
\newblock \showarticletitle{Clipper: A low-latency online prediction serving
  system}. In \bibinfo{booktitle}{\emph{14th $\{$USENIX$\}$ Symposium on
  Networked Systems Design and Implementation ($\{$NSDI$\}$ 17)}}.
  \bibinfo{publisher}{USENIX Association}, \bibinfo{address}{CA, United
  States}, \bibinfo{pages}{613--627}.
\newblock


\bibitem[\protect\citeauthoryear{Falcon}{Falcon}{2019}]%
        {falcon2019pytorch}
\bibfield{author}{\bibinfo{person}{et~al. Falcon, WA}.}
  \bibinfo{year}{2019}\natexlab{}.
\newblock \showarticletitle{PyTorch Lightning}.
\newblock \bibinfo{journal}{\emph{GitHub. Note:
  https://github.com/PyTorchLightning/pytorch-lightning}}  \bibinfo{volume}{3}
  (\bibinfo{year}{2019}).
\newblock


\bibitem[\protect\citeauthoryear{Haldar, Abdool, Ramanathan, Xu, Yang, Duan,
  Zhang, Barrow-Williams, Turnbull, Collins, et~al\mbox{.}}{Haldar
  et~al\mbox{.}}{2019}]%
        {haldar2019applying}
\bibfield{author}{\bibinfo{person}{Malay Haldar}, \bibinfo{person}{Mustafa
  Abdool}, \bibinfo{person}{Prashant Ramanathan}, \bibinfo{person}{Tao Xu},
  \bibinfo{person}{Shulin Yang}, \bibinfo{person}{Huizhong Duan},
  \bibinfo{person}{Qing Zhang}, \bibinfo{person}{Nick Barrow-Williams},
  \bibinfo{person}{Bradley~C Turnbull}, \bibinfo{person}{Brendan~M Collins},
  {et~al\mbox{.}}} \bibinfo{year}{2019}\natexlab{}.
\newblock \showarticletitle{Applying deep learning to Airbnb search}. In
  \bibinfo{booktitle}{\emph{Proceedings of the 25th ACM SIGKDD International
  Conference on Knowledge Discovery \& Data Mining}}.
  \bibinfo{publisher}{Association for Computing Machinery},
  \bibinfo{address}{NY, USA}, \bibinfo{pages}{1927--1935}.
\newblock


\bibitem[\protect\citeauthoryear{Hinton, Vinyals, and Dean}{Hinton
  et~al\mbox{.}}{2015}]%
        {hinton2015distilling}
\bibfield{author}{\bibinfo{person}{Geoffrey Hinton}, \bibinfo{person}{Oriol
  Vinyals}, {and} \bibinfo{person}{Jeffrey Dean}.}
  \bibinfo{year}{2015}\natexlab{}.
\newblock \showarticletitle{Distilling the Knowledge in a Neural Network}. In
  \bibinfo{booktitle}{\emph{NIPS Deep Learning and Representation Learning
  Workshop}}.
\newblock
\urldef\tempurl%
\url{http://arxiv.org/abs/1503.02531}
\showURL{%
\tempurl}


\bibitem[\protect\citeauthoryear{Li and Zaki}{Li and Zaki}{2020}]%
        {li2020reciptor}
\bibfield{author}{\bibinfo{person}{Diya Li} {and} \bibinfo{person}{Mohammed~J
  Zaki}.} \bibinfo{year}{2020}\natexlab{}.
\newblock \showarticletitle{Reciptor: An effective pretrained model for recipe
  representation learning}. In \bibinfo{booktitle}{\emph{Proceedings of the
  26th ACM SIGKDD International Conference on Knowledge Discovery \& Data
  Mining}}. \bibinfo{publisher}{Association for Computing Machinery},
  \bibinfo{address}{NY, USA}, \bibinfo{pages}{1719--1727}.
\newblock


\bibitem[\protect\citeauthoryear{Long, Cao, Wang, and Jordan}{Long
  et~al\mbox{.}}{2015}]%
        {long2015learning}
\bibfield{author}{\bibinfo{person}{Mingsheng Long}, \bibinfo{person}{Yue Cao},
  \bibinfo{person}{Jianmin Wang}, {and} \bibinfo{person}{Michael Jordan}.}
  \bibinfo{year}{2015}\natexlab{}.
\newblock \showarticletitle{Learning transferable features with deep adaptation
  networks}. In \bibinfo{booktitle}{\emph{Proceedings of the 32nd International
  Conference on International Conference on Machine Learning}}. PMLR,
  \bibinfo{publisher}{JMLR.org}, \bibinfo{pages}{97--105}.
\newblock


\bibitem[\protect\citeauthoryear{Luo, Zhang, Wen, and Zhang}{Luo
  et~al\mbox{.}}{2019}]%
        {luo2019resumegan}
\bibfield{author}{\bibinfo{person}{Yong Luo}, \bibinfo{person}{Huaizheng
  Zhang}, \bibinfo{person}{Yonggang Wen}, {and} \bibinfo{person}{Xinwen
  Zhang}.} \bibinfo{year}{2019}\natexlab{}.
\newblock \showarticletitle{ResumeGAN: An Optimized Deep Representation
  Learning Framework for Talent-Job Fit via Adversarial Learning}. In
  \bibinfo{booktitle}{\emph{Proceedings of the 28th ACM International
  Conference on Information and Knowledge Management}}.
  \bibinfo{pages}{1101--1110}.
\newblock


\bibitem[\protect\citeauthoryear{Michel, Holbrook, Grosser, Strobelt, and
  Shah}{Michel et~al\mbox{.}}{[n.d.]}]%
        {ennui}
\bibfield{author}{\bibinfo{person}{Jesse Michel}, \bibinfo{person}{Zack
  Holbrook}, \bibinfo{person}{Stefan Grosser}, \bibinfo{person}{Hendrik
  Strobelt}, {and} \bibinfo{person}{Rikhav Shah}.}
  \bibinfo{year}{[n.d.]}\natexlab{}.
\newblock \bibinfo{title}{ENNUI ~ Elegant Neural Network User Interface ~}.
\newblock
\newblock
\urldef\tempurl%
\url{https://math.mit.edu/ennui/}
\showURL{%
\tempurl}
\newblock
\shownote{Accessed: 2021-02-26.}


\bibitem[\protect\citeauthoryear{NVIDIA}{NVIDIA}{2020}]%
        {tlt}
\bibfield{author}{\bibinfo{person}{NVIDIA}.} \bibinfo{year}{2020}\natexlab{}.
\newblock \bibinfo{title}{Transfer Learning Toolkit.}
\newblock
\newblock
\urldef\tempurl%
\url{https://developer.nvidia.com/transfer-learning-toolkit}
\showURL{%
\tempurl}
\newblock
\shownote{Accessed: 2021-03-15.}


\bibitem[\protect\citeauthoryear{Pan and Yang}{Pan and Yang}{2009}]%
        {pan2009survey}
\bibfield{author}{\bibinfo{person}{Sinno~Jialin Pan} {and}
  \bibinfo{person}{Qiang Yang}.} \bibinfo{year}{2009}\natexlab{}.
\newblock \showarticletitle{A survey on transfer learning}.
\newblock \bibinfo{journal}{\emph{IEEE Transactions on knowledge and data
  engineering}} \bibinfo{volume}{22}, \bibinfo{number}{10}
  (\bibinfo{year}{2009}), \bibinfo{pages}{1345--1359}.
\newblock


\bibitem[\protect\citeauthoryear{Pardoe and Stone}{Pardoe and Stone}{2010}]%
        {pardoe2010boosting}
\bibfield{author}{\bibinfo{person}{David Pardoe} {and} \bibinfo{person}{Peter
  Stone}.} \bibinfo{year}{2010}\natexlab{}.
\newblock \showarticletitle{Boosting for regression transfer}. In
  \bibinfo{booktitle}{\emph{Proceedings of the 27th International Conference on
  International Conference on Machine Learning}}.
  \bibinfo{publisher}{Omnipress}, \bibinfo{address}{WI, United States},
  \bibinfo{pages}{863–--870}.
\newblock


\bibitem[\protect\citeauthoryear{Qiu, Li, Pan, Wang, Chen, Li, Gao, Huang, Li,
  Yang, et~al\mbox{.}}{Qiu et~al\mbox{.}}{2020}]%
        {qiu2020easytransfer}
\bibfield{author}{\bibinfo{person}{Minghui Qiu}, \bibinfo{person}{Peng Li},
  \bibinfo{person}{Hanjie Pan}, \bibinfo{person}{Chengyu Wang},
  \bibinfo{person}{Cen Chen}, \bibinfo{person}{Yaliang Li},
  \bibinfo{person}{Dehong Gao}, \bibinfo{person}{Jun Huang},
  \bibinfo{person}{Yong Li}, \bibinfo{person}{Jun Yang}, {et~al\mbox{.}}}
  \bibinfo{year}{2020}\natexlab{}.
\newblock \showarticletitle{EasyTransfer--A Simple and Scalable Deep Transfer
  Learning Platform for NLP Applications}.
\newblock \bibinfo{journal}{\emph{arXiv preprint arXiv:2011.09463}}
  (\bibinfo{year}{2020}).
\newblock


\bibitem[\protect\citeauthoryear{Roeder}{Roeder}{[n.d.]}]%
        {netron}
\bibfield{author}{\bibinfo{person}{Lutz Roeder}.}
  \bibinfo{year}{[n.d.]}\natexlab{}.
\newblock \bibinfo{title}{Visualizer for neural network, deep learning, and
  machine learning models}.
\newblock
\newblock
\urldef\tempurl%
\url{https://github.com/lutzroeder/netron}
\showURL{%
\tempurl}
\newblock
\shownote{Accessed: 2021-02-26.}


\bibitem[\protect\citeauthoryear{Snoek, Larochelle, and Adams}{Snoek
  et~al\mbox{.}}{2012}]%
        {snoek2012practical}
\bibfield{author}{\bibinfo{person}{Jasper Snoek}, \bibinfo{person}{Hugo
  Larochelle}, {and} \bibinfo{person}{Ryan~P Adams}.}
  \bibinfo{year}{2012}\natexlab{}.
\newblock \showarticletitle{Practical bayesian optimization of machine learning
  algorithms}. In \bibinfo{booktitle}{\emph{Proceedings of the 25th
  International Conference on Neural Information Processing Systems - Volume
  2}}. \bibinfo{publisher}{Curran Associates Inc.}, \bibinfo{address}{NY,
  United States}, \bibinfo{pages}{2951–--2959}.
\newblock


\bibitem[\protect\citeauthoryear{Sony}{Sony}{[n.d.]}]%
        {nnc}
\bibfield{author}{\bibinfo{person}{Sony}.} \bibinfo{year}{[n.d.]}\natexlab{}.
\newblock \bibinfo{title}{Neural Network Console - Not just train and evaluate.
  You can design neural networks with fast and intuitive GUI}.
\newblock
\newblock
\urldef\tempurl%
\url{https://dl.sony.com/}
\showURL{%
\tempurl}


\bibitem[\protect\citeauthoryear{Stojnic, Taylor, Kardas, Kerkez, Viaud,
  Saravia, and Cucurull}{Stojnic et~al\mbox{.}}{[n.d.]}]%
        {paperswithcode}
\bibfield{author}{\bibinfo{person}{Robert Stojnic}, \bibinfo{person}{Ross
  Taylor}, \bibinfo{person}{Marcin Kardas}, \bibinfo{person}{Viktor Kerkez},
  \bibinfo{person}{Ludovic Viaud}, \bibinfo{person}{Elvis Saravia}, {and}
  \bibinfo{person}{Guillem Cucurull}.} \bibinfo{year}{[n.d.]}\natexlab{}.
\newblock \bibinfo{title}{Papers with Code - The latest in Machine Learning}.
\newblock
\newblock
\urldef\tempurl%
\url{https://paperswithcode.com/}
\showURL{%
\tempurl}
\newblock
\shownote{Accessed: 2021-05-01.}


\bibitem[\protect\citeauthoryear{Xu, Li, Cui, Huang, Wei, and Zhou}{Xu
  et~al\mbox{.}}{2020}]%
        {xu2020layoutlm}
\bibfield{author}{\bibinfo{person}{Yiheng Xu}, \bibinfo{person}{Minghao Li},
  \bibinfo{person}{Lei Cui}, \bibinfo{person}{Shaohan Huang},
  \bibinfo{person}{Furu Wei}, {and} \bibinfo{person}{Ming Zhou}.}
  \bibinfo{year}{2020}\natexlab{}.
\newblock \showarticletitle{Layoutlm: Pre-training of text and layout for
  document image understanding}. In \bibinfo{booktitle}{\emph{Proceedings of
  the 26th ACM SIGKDD International Conference on Knowledge Discovery \& Data
  Mining}}. \bibinfo{publisher}{Association for Computing Machinery},
  \bibinfo{address}{NY, USA}, \bibinfo{pages}{1192--1200}.
\newblock


\bibitem[\protect\citeauthoryear{You, Liu, Long, and Wang}{You
  et~al\mbox{.}}{2021}]%
        {you_logme:_2021}
\bibfield{author}{\bibinfo{person}{Kaichao You}, \bibinfo{person}{Yong Liu},
  \bibinfo{person}{Mingsheng Long}, {and} \bibinfo{person}{Jianmin Wang}.}
  \bibinfo{year}{2021}\natexlab{}.
\newblock \showarticletitle{LogME: Practical Assessment of Pre-trained Models
  for Transfer Learning}.
\newblock \bibinfo{journal}{\emph{arxiv}}  \bibinfo{volume}{abs/2102.11005}
  (\bibinfo{year}{2021}).
\newblock
\urldef\tempurl%
\url{https://arxiv.org/abs/2102.11005}
\showURL{%
\tempurl}


\bibitem[\protect\citeauthoryear{Zhang, Huang, Wen, Yin, and Guan}{Zhang
  et~al\mbox{.}}{2020a}]%
        {zhang2020no}
\bibfield{author}{\bibinfo{person}{Huaizheng Zhang}, \bibinfo{person}{Yizheng
  Huang}, \bibinfo{person}{Yonggang Wen}, \bibinfo{person}{Jianxiong Yin},
  {and} \bibinfo{person}{Kyle Guan}.} \bibinfo{year}{2020}\natexlab{a}.
\newblock \showarticletitle{No more 996: Understanding Deep Learning Inference
  Serving with an Automatic Benchmarking system}.
\newblock \bibinfo{journal}{\emph{arXiv preprint arXiv:2011.02327}}
  (\bibinfo{year}{2020}).
\newblock


\bibitem[\protect\citeauthoryear{Zhang, Li, Ai, Luo, Wen, Jin, and Ta}{Zhang
  et~al\mbox{.}}{2020b}]%
        {hysia}
\bibfield{author}{\bibinfo{person}{Huaizheng Zhang}, \bibinfo{person}{Yuanming
  Li}, \bibinfo{person}{Qiming Ai}, \bibinfo{person}{Yong Luo},
  \bibinfo{person}{Yonggang Wen}, \bibinfo{person}{Yichao Jin}, {and}
  \bibinfo{person}{Nguyen Binh~Duong Ta}.} \bibinfo{year}{2020}\natexlab{b}.
\newblock \showarticletitle{Hysia: Serving DNN-Based Video-to-Retail
  Applications in Cloud}. In \bibinfo{booktitle}{\emph{Proceedings of the 28th
  ACM International Conference on Multimedia}}. \bibinfo{publisher}{Association
  for Computing Machinery}, \bibinfo{address}{NY, USA},
  \bibinfo{pages}{4457--4460}.
\newblock


\bibitem[\protect\citeauthoryear{Zhang, Li, Huang, Wen, Yin, and Guan}{Zhang
  et~al\mbox{.}}{2020c}]%
        {zhang2020mlmodelci}
\bibfield{author}{\bibinfo{person}{Huaizheng Zhang}, \bibinfo{person}{Yuanming
  Li}, \bibinfo{person}{Yizheng Huang}, \bibinfo{person}{Yonggang Wen},
  \bibinfo{person}{Jianxiong Yin}, {and} \bibinfo{person}{Kyle Guan}.}
  \bibinfo{year}{2020}\natexlab{c}.
\newblock \showarticletitle{Mlmodelci: An automatic cloud platform for
  efficient mlaas}. In \bibinfo{booktitle}{\emph{Proceedings of the 28th ACM
  International Conference on Multimedia}}. \bibinfo{publisher}{Association for
  Computing Machinery}, \bibinfo{address}{NY, USA},
  \bibinfo{pages}{4453--4456}.
\newblock


\end{thebibliography}

\end{document}